\documentclass[twoside,english]{elsarticle}
\usepackage[T1]{fontenc}
\usepackage[latin9]{inputenc}
\usepackage{array}
\usepackage{textcomp}
\usepackage{url}
\usepackage{multirow}
\usepackage{amstext}
\usepackage{graphicx}

\makeatletter

\providecommand{\tabularnewline}{\\}

\journal{HardwareX}

\usepackage{lineno}

\makeatother

\usepackage{babel}
\begin{document}

\begin{frontmatter}{}

\title{A cost effective and reliable environment monitoring system for HPC
applications}

\author[rvt]{Peter-Bernd~Otte\corref{cor1}}

\ead{p.otte@him.uni-mainz.de}

\ead[url]{https://him.uni-mainz.de/scientific-computing}

\author[rvt]{Dalibor~Djukanovic}

\cortext[cor1]{Corresponding author}

\address[rvt]{Helmholtz-Institut Mainz, Staudinger Weg 18, 55128 Mainz, Germany}

\address{}
\begin{abstract}
We present a slow control system to gather all relevant environment
information necessary to effectively and reliably run an HPC (High
Performance Computing) system at a high value over price ratio. The
scalable and reliable overall concept is presented as well as a newly
developed hardware device for sensor read out. This device incorporates
a Raspberry Pi, an Arduino and PoE (Power over Ethernet) functionality
in a compact form factor. The system is in use at the 2\,PFLOPS cluster
of the Johannes Gutenberg-University and Helmholtz-Institute in Mainz.
\end{abstract}

\end{frontmatter}{}


\section{Introduction}

HPC systems are typically expensive large-scale devices in research.
Consequently, monitoring is desirable for securing data and values
(hardware). Furthermore, such environment monitoring also enables
the optimization of operating parameters to increase efficiency (equivalent
of minimizing the Power Usage Effectiveness, PUE) and thus a reduction
of the carbon dioxide footprint.

In the HPC environment, hundreds of data points are collected usually
at geographically widely scattered locations within several venues.
The typical readout rate is in the order of about 1\,Hz.

During the development phase, we aimed for a very high degree of reliability
while at the same time simplifying the process of installation and
operation. In addition, the use of widely used open source components
(Raspberry Pi, Arduino) increases familiarity with the system components,
i.e. enables fast installation, low cost of solution roll out and
good debugging possibilities. The overall system can be crucial for
operating any HPC system efficiently and safely.

Our environment monitoring system setup currently reads temperatures
of water and air, humidity levels, water flow meters and water leakage
sensors. The data is sent to a data collecting server running Ganglia
\cite{key-Ganglia} which enables time series monitoring as well as
notifications in case of alarming conditions.

The problem of monitoring environmental data is not restricted to
computing centers, where a lot of, in most parts redundant, effort
is put into the solution of problems of this kind at various sites,
both using commercial and custom made products. This work aims at
documenting an all, i.e. including hardware design, open source solution
to measuring and processing environmental data, which is flexible
enough to incorporate various sensors while still being easy to implement.

\section{Design}

\subsection{Overview}

\begin{figure}
\includegraphics[width=1\textwidth]{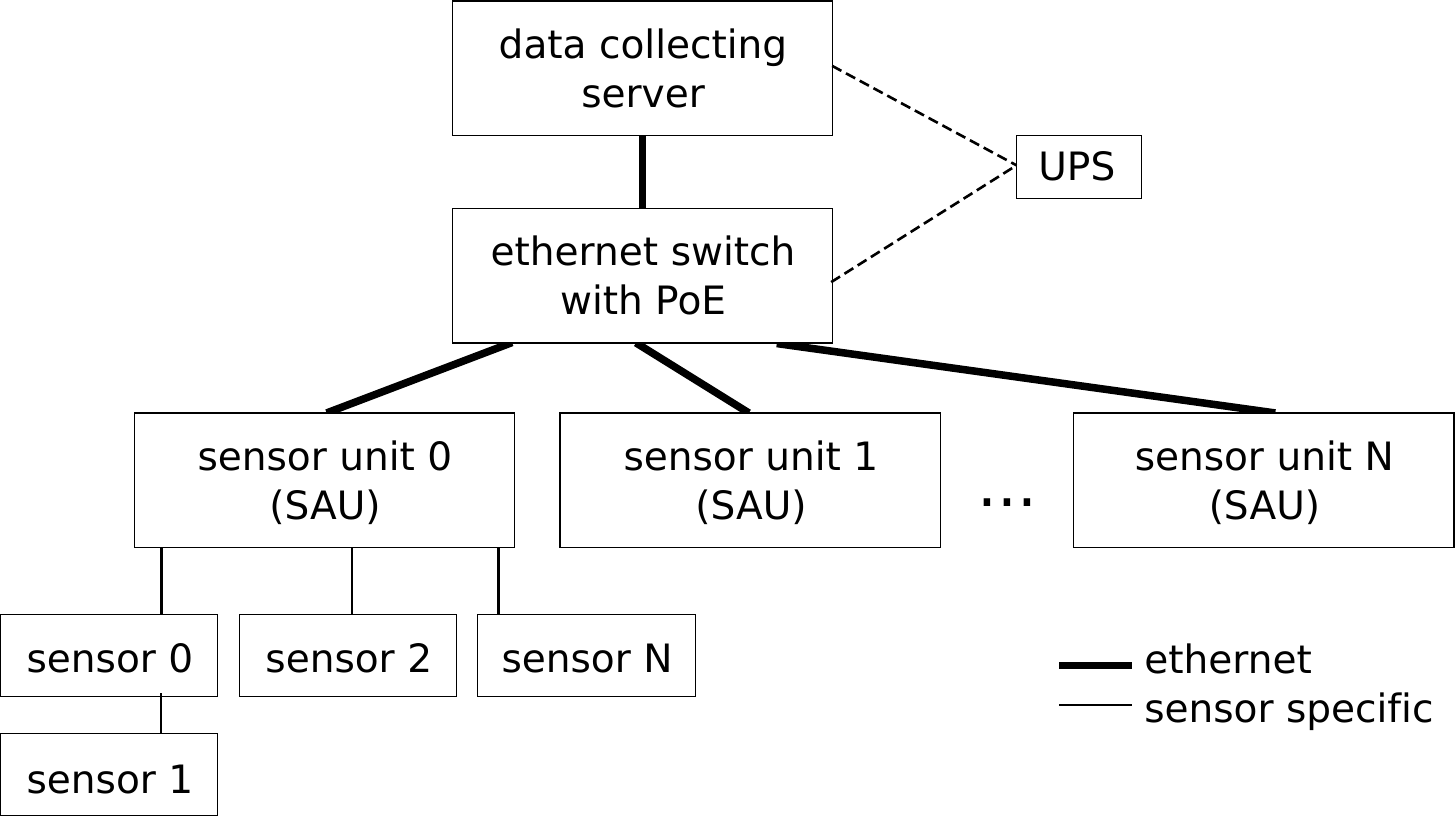}

\caption{Overview of overall setup\label{fig:Overview}}

\end{figure}

Our System consists of the following main elements: sensors, sensor
aggregation unit (SAU) and a data collecting server \textendash{}
see the overview image \ref{fig:Overview}.

Further necessary parts are a Power over Ethernet (PoE, IEEE 802.3af)
enabled fast Ethernet switch (100\,Mbit/s) as well as a single uninterruptible
power supply (UPS) to overcome power outages or discontinuities. Due
to the PoE enabled SAUs, the complete monitoring system down to the
smallest sensor is UPS protected. In addition, careful selection of
the Ethernet switch adds the functionality to remotely power cycle
every sensor and sensor aggregation units out of the box.

\subsection{Monitoring Network }

Environmental sensors are connected to the sensor aggregation unit,
get read out and send via an Ethernet link to the central collecting
server. While the radius of Ethernet is specified for up to 100 meters,
the sensor radius connected to a single SAU is limited to about 10
meters due to the usage of inexpensive RJ12 connectors, Y splitters
and standard telephone cables. In a real environment, both limitations
are negligible. As a side effect, the RJ12 ensures that no RJ45 network
cable from the remaining HPC system gets plugged into the sensor network
by mistake. The communication over the telephone cables are sensor
specific and reach from static analog voltages to digital bus systems
(at frequencies below 1\,MHz).

\subsection{Sensor Aggregation Unit}

\begin{figure}
\includegraphics[angle=180,width=0.49\textwidth]{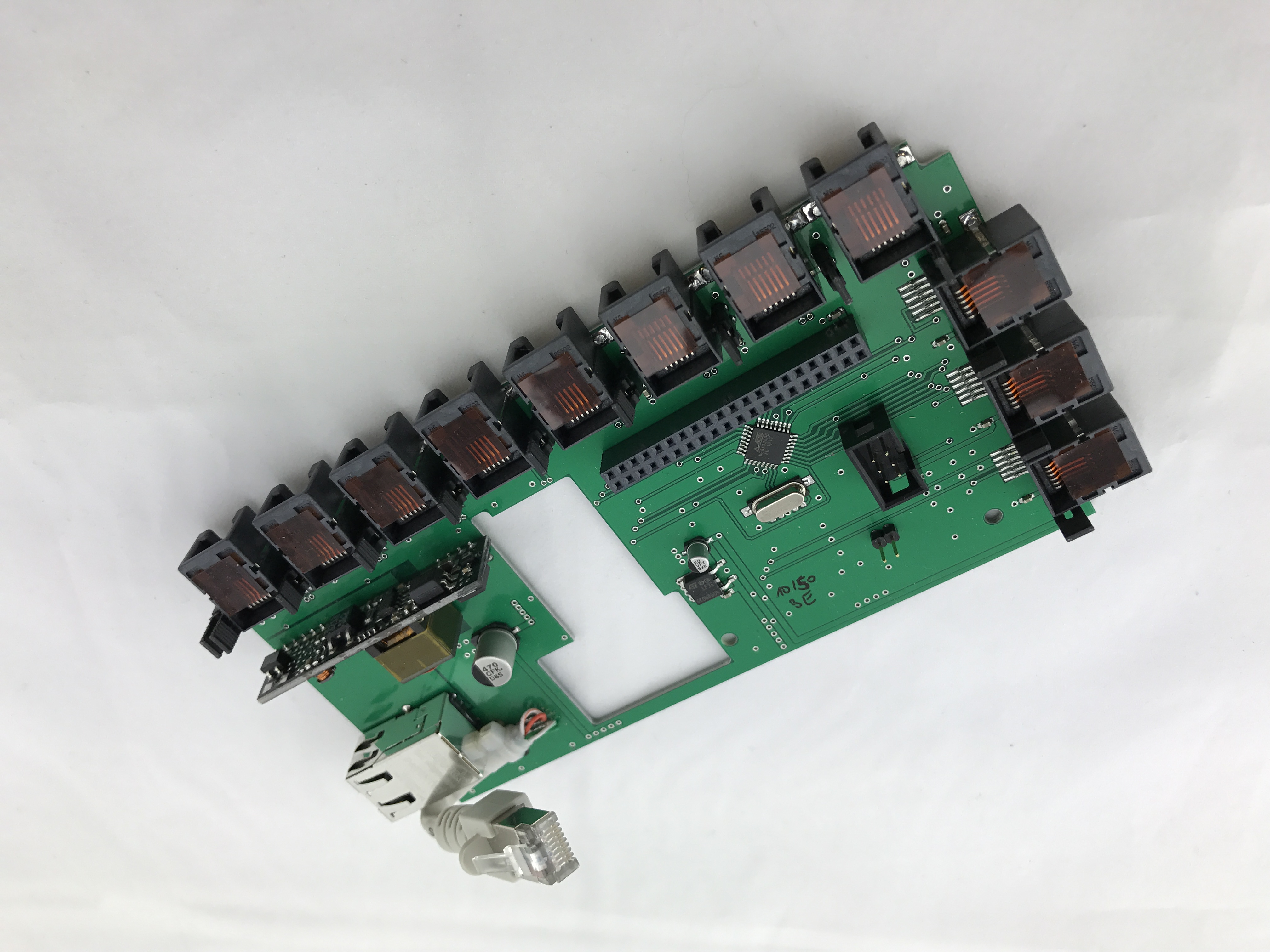}
\includegraphics[angle=180,width=0.49\textwidth]{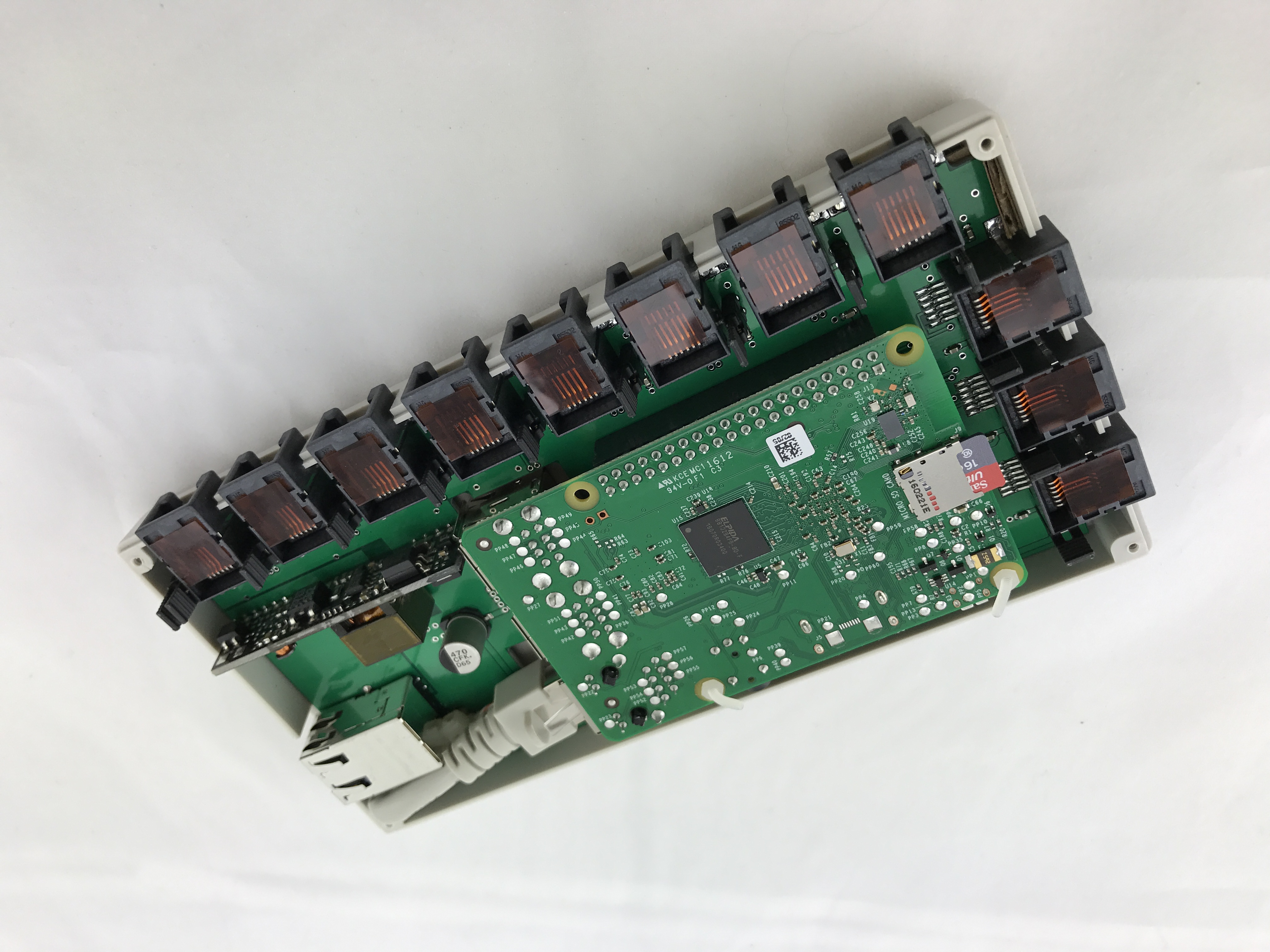}

\includegraphics[angle=180,width=0.49\textwidth]{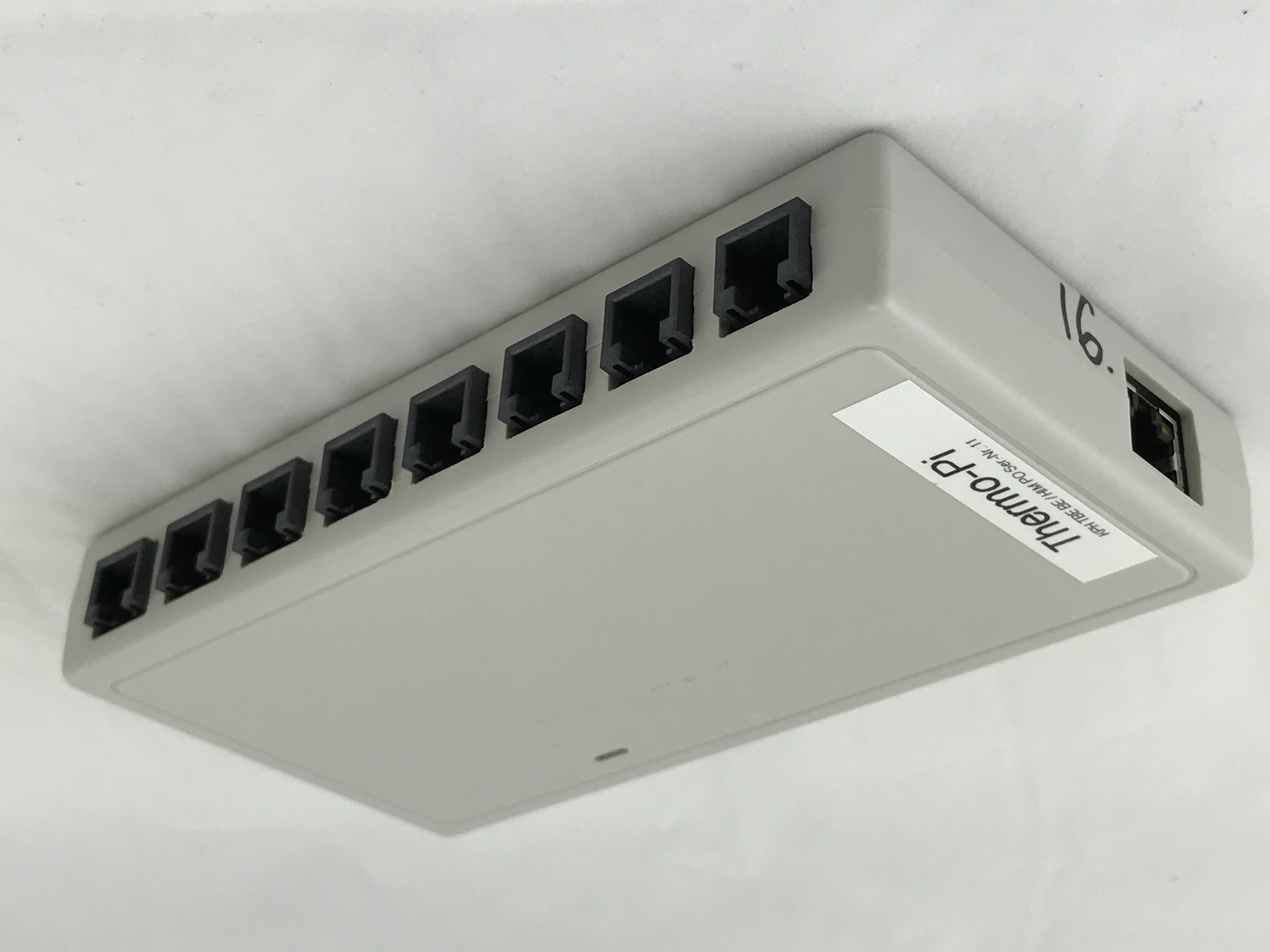}
\includegraphics[angle=180,width=0.49\textwidth]{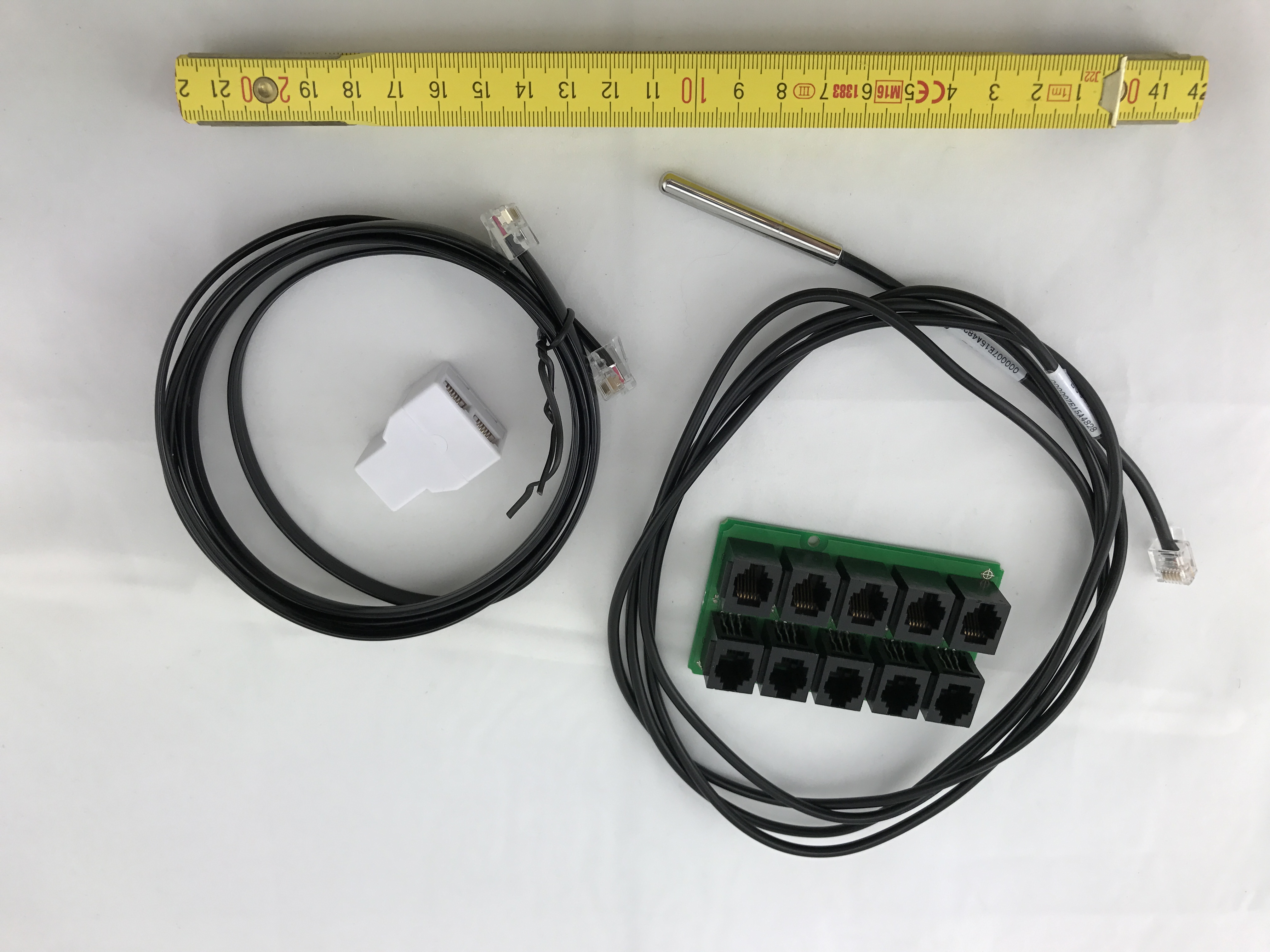}

\caption{Sensor Aggregation Unit (SAU) and sensor cables\label{fig:Sensor-Aggregation-Unit}}
\end{figure}

A sensor aggregation unit (SAU) consists of a mainboard and a Raspberry
Pi as an add-on (see Fig. \ref{fig:Sensor-Aggregation-Unit}).
The mainboard provides the PoE functionality (IEEE 802.3af), an Arduino
(Atmel ATMega 328PB) and bus driver (OneWire and I2C) functionality.
The connection between the Arduino and the Raspberry Pi is done via
its 40pin connector. Outgoing connections are only provided by the
mainboard, these are the PoE enabled network connector and eleven
RJ12 sensor ports. In total, these design decisions make the installation
very compact and handy.

The decision to introduce an Arduino and to not connect the sensors
directly to the Raspberry Pi was made for reasons of reliability and
flexibility (due to program dependent pin usage). A serial connection
together with a reset line links the Arduino with the Raspberry Pi.
The reset line enables not only a reset of the Arduino but also in-system
flashing. To enable this functionality, the optiboot bootloader has
been preloaded via the ISP port on the mainboard. According to the
design of the mainboard (16\,MHz external low power crystal) low
and high fuse bits have been set to 0xde {[}4{]}.

Furthermore tests revealed an unstable behavior for a setup of four
DS18B20 directly connected to the pins of a Raspberry Pi 2B with the
latest (March 2017) Raspbian software release. To resolve these problems,
the Raspberry Pi had to be power cycled because a simple reboot was
not sufficient.

To further strengthen reliability the 5V line (generated by the PoE
sub-module) powering the Raspberry Pi is not accessible from outside
the sensor aggregation unit. Only the 3.3V derived by a low-dropout
regulator (LDO) from it is available at the sensor connectors. Shortcuts
on one of the sensor connectors will only affect the Arduino part,
but will not bring the Raspberry Pi into an unstable state. As soon
as the shortcut is resolved, the activated Brown-out Detection (BOD)
(extended fuse bits = 0xfd) of the ATMega helps recovering immediately
to a save state of operation.

Further benefits of the ATMega chip are its realtime capability and
analog inputs to read out even more diverse sensors compared to the
Raspberry Pi only solution.

The pin assignment of a RJ12 sensor port is listed in Tab. \ref{tab:Pin-assignment-for}.
They all share a common I2C bus (with optional internal 2.2\,$\text{k}\Omega$
pull-ups, according to {[}5{]}) and have two additional I/O ports.
One of it is enabled to act as an OneWire master, by activating the
individual internal 4.7\,$\text{k}\Omega$ pull-up. The second provides
an analog I/O port for ports 1-6 and digital I/O for the remaining
ports to read out an even larger selection of different sensor types.
\begin{table}
\begin{tabular}{|c|c|}
\hline
Pin \# & Description\tabularnewline
\hline
\hline
1 & 3.3 VCC\tabularnewline
\hline
2 & Analog I/O (ports 1-6) or Digital I/O (ports 7-11) \tabularnewline
\hline
3 & Digital I/O (with optional individual pull-up for OneWire data)\tabularnewline
\hline
4 & I2C data (common for all ports, optional pull-up)\tabularnewline
\hline
5 & GND\tabularnewline
\hline
6 & I2C clock (common for all ports, optional pull-up)\tabularnewline
\hline
\end{tabular}

\caption{Pin assignment for the RJ12 sensor ports\label{tab:Pin-assignment-for}}

\end{table}

\begin{figure}
\includegraphics[width=1\textwidth]{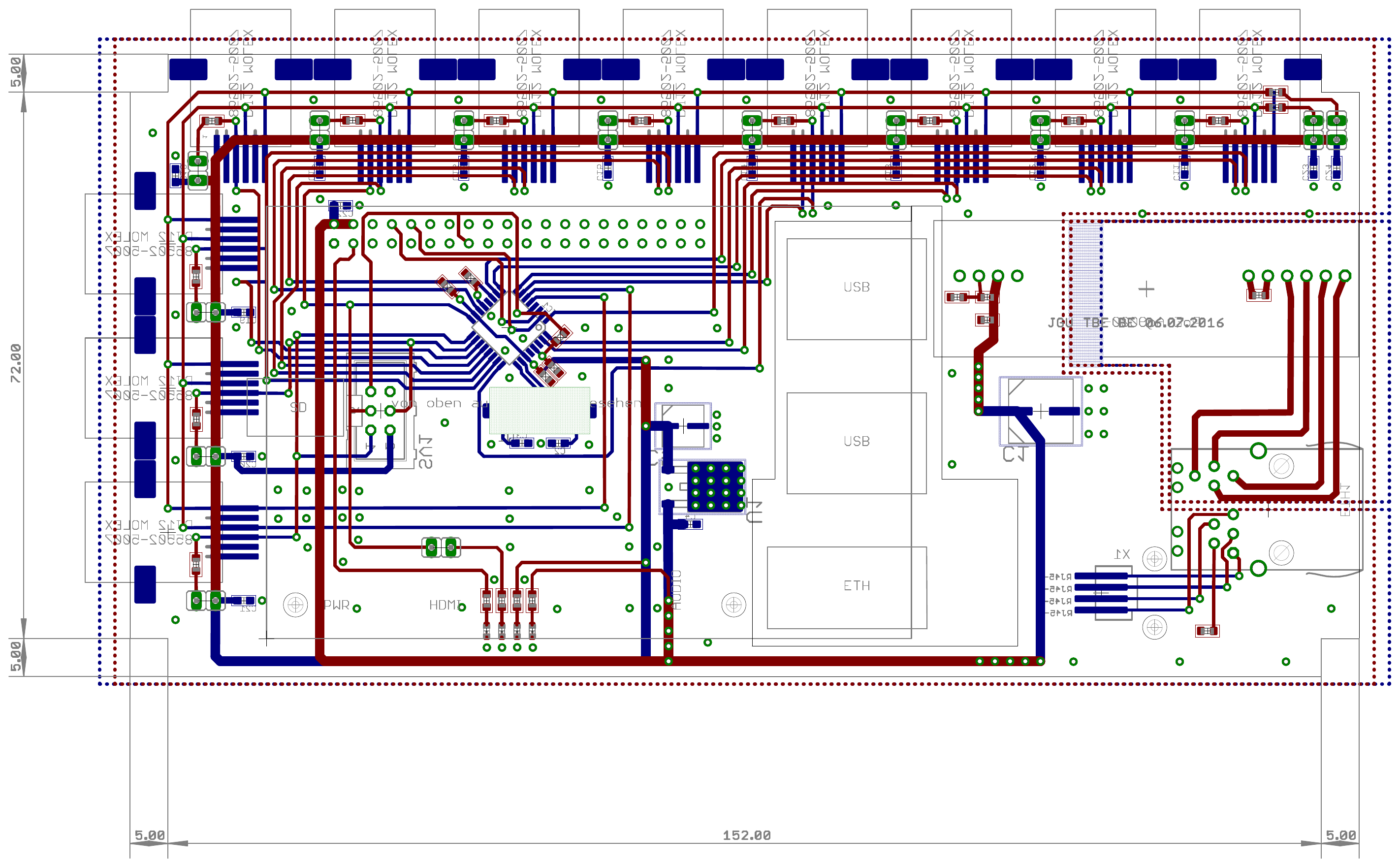}

\includegraphics[width=1\textwidth]{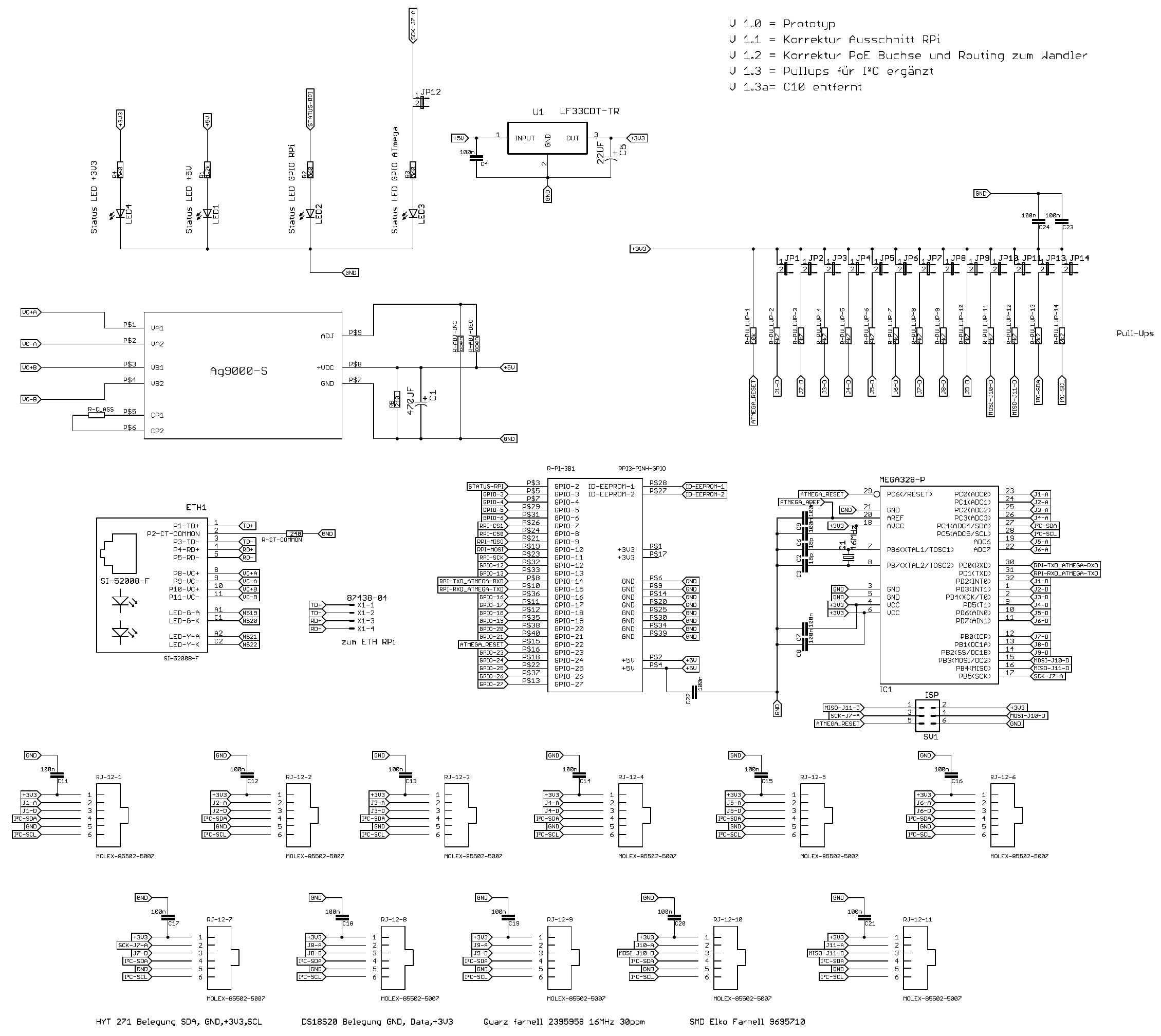}

\caption{Electronics Layout\label{fig:Electronics-Layout}}

\end{figure}

The sensor aggregation unit schematics and the PCB layout are shown
in Fig. \ref{fig:Electronics-Layout}.

\subsection{Sensors Recommendations }

\begin{figure}
\begin{centering}
\includegraphics[width=0.6\textwidth]{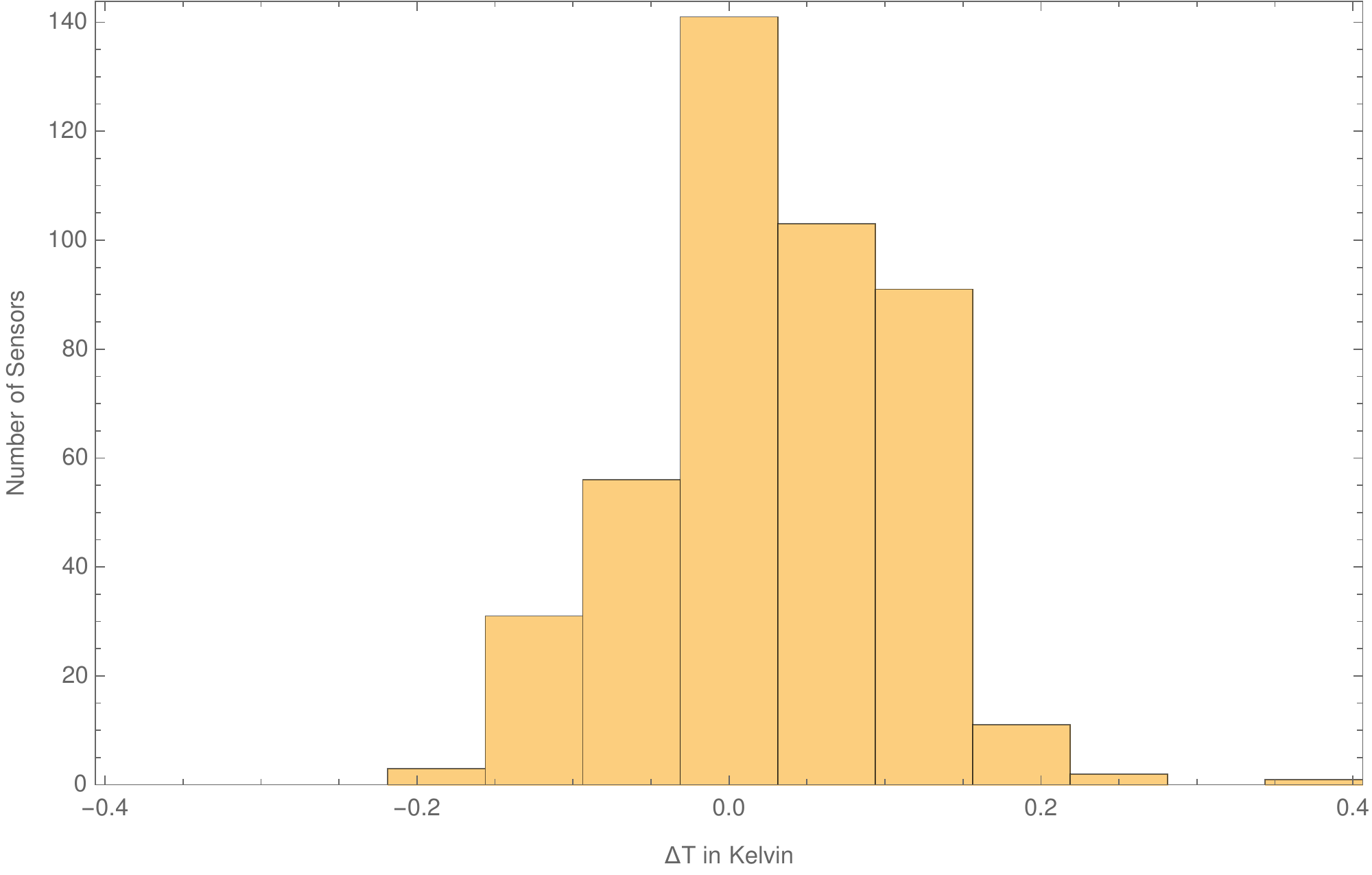}
\par\end{centering}
\caption{Miscalibration of 502 DS18B20 sensors at 20\,\textdegree C. \cite{key-Sergey}\label{fig:Miscalibration-of-502-DS18B20}}

\end{figure}

Among several sensors tested, the following were selected and are
used in our setup. They fulfill a reasonable mix of performance and
stability in comparison to their price:
\begin{itemize}
\item air / water temperature: Maxim DS18B20, sealed in stainless steel
pipe, connection type: OneWire, response time: in still air $\tau\approx90\text{\thinspace sec}$,
accuracy $\Delta T=\pm0.5\thinspace\text{K}$ \cite{key-DS18B20-Datasheet}
\item air temperature and humidity: IST HYT-271, connection type: I2C, response
time: in still air $\tau\approx180\text{\thinspace sec}$, with air-flow
$\tau\approx4\text{\thinspace sec}$, accuracy $\Delta T=\pm0.2\thinspace\text{K}$
\cite{key-HYT271-Datasheet}
\item water flow: water meter with reed contact as binary input
\item air pressure: Bosch BME280 (together with temperature and humidity),
connection type: I2C, response time: in still air $\tau\approx270\text{\thinspace sec}$,
with air-flow $\tau\approx1\text{\thinspace sec}$, accuracy $\Delta T=\pm1.0\thinspace\text{K}$
\cite{key-BoschBME280}
\end{itemize}
Some remarks for two of the above sensors: First, we custom ordered
the OneWire temperature sensor from Maxim in a water-resistant packaging
and a RJ12 connector matching our pin assignment and cable length.
This makes installation as an air or water temperature sensor quick
and easy. To increase the absolute precision of the sensor, we calibrated
the offset individually before installation at 20\,\textdegree C,
see the result for 502 sensors in Fig. \ref{fig:Miscalibration-of-502-DS18B20}.
This recalibration is done in the same units as the sensor outputs
its data: degree Celsius.

Secondly, the Bosch BME280 needs in general a recalibration over the
complete temperature range. With the given individual sensor calibration
constants, the temperature values do have a miscalibration in the
order of $\pm1\thinspace\text{K}$ (compare Tab. \ref{tab:Recalibration-of-BME280}).
This calibration fault increases further due to self heating and can
be reduced when putting the sensor extensively into sleep mode. As
a follow-up error, the temperature readings affect the humidity and
pressure values. After recalibration, temperature readings operate
reasonably stable (better than 0.1\,K deviation over the complete
range).

In the following, we describe the recalibration in a few steps: The
sensor of interest is put together with a precise reference into a
climate chamber and is exposed to temperatures between -40\,\textdegree C
and 60\textdegree C. The temperature is ramped up and down with less
then 0.2 \textdegree C/min to allow for assimilation.

The data sheet \cite{key-BoschBME280} suggests a quadratic polynomial
to calculate the real temperature $T$ {[}\textdegree C{]} with the
measured raw-value $T_{\text{raw}}$
\[
T=c_{0}+c_{1}T_{\text{raw}}+c_{2}T_{\text{raw}}^{2}
\]
 and
\begin{eqnarray*}
c_{0} & = & -\frac{4}{5\cdot2^{22}}\left(d_{1}d_{2}-\frac{d_{1}^{2}d_{3}}{2^{16}}\right)\\
c_{1} & = & \frac{4}{5\cdot2^{26}}\left(d_{2}-\frac{d_{1}d_{3}}{2^{15}}\right)\\
c_{2} & = & \frac{4}{5\cdot2^{46}}d_{3}
\end{eqnarray*}
where the constants $d_{i}$ have been calibrated in the factory according
to the device. Fitting the measured data and choosing the positive
solution provides a new set of high precision constants:

\begin{eqnarray*}
d_{1} & = & -\frac{c_{1}+\sqrt{c_{1}^{2}-4c_{0}c_{2}}}{32c_{2}}\\
d_{2} & = & \frac{5\cdot2^{26}}{4}\sqrt{c_{1}^{2}-4c_{0}c_{2}}\\
d_{3} & = & \frac{5\cdot2^{46}}{4}c_{2}
\end{eqnarray*}

Example values for four sensors, before and after recalibration are
given in Tab. \ref{tab:Recalibration-of-BME280}.

\begin{table}
\begin{tabular}{|c|c|c|c|c|c|c||c|}
\hline
\multirow{2}{*}{sensor} & \multicolumn{3}{c|}{factory calibration} & \multicolumn{3}{c||}{new calibration} & deviation\tabularnewline
\cline{2-8}
 & $d_{1}$ & $d_{2}$ & $d_{3}$ & $d_{1}$ & $d_{2}$ & $d_{3}$ & -40..60\,\textdegree C\tabularnewline
\hline
\hline
1 & 28205 & 28205 & 50 & 28469 & 26034 & 753.63 & -0.6 .. -1.8\tabularnewline
\hline
2 & 28498 & 26766 & 50 & 30462 & 23501 & 2846.84 & -2 .. -14\tabularnewline
\hline
3 & 28222 & 26702 & 50 & 28172 & 26073 & -388.33 & 1.2 .. -1.3\tabularnewline
\hline
4 & 28266 & 26340 & 50 & 28304 & 26409 & 299.61 & -0.3 .. 0.0\tabularnewline
\hline
\end{tabular}

\caption{Recalibration of BME280 sensors, $d_{1}$ is dimensionless, $d_{2}$
and $d_{3}$ are in \textdegree C.\label{tab:Recalibration-of-BME280}}

\end{table}

\subsection{Cabling and its Limitations }

In this paragraph, we want to shortly describe some best practice
hints for the sensor connecting cables as well as its limitations.
Generally speaking, the maximum cable length and its network topology
strongly depends on the electrical characteristics of the cables,
passive distributor panels, number of sensors and bus type. To simplify
the setup, we restricted our OneWire bus to a maximum radius of 10
Meter and not more than 15 sensors as we identified that problems
occurred with our cabling hardware at about a 50\,m radius. There
is an interesting feature of the OneWire bus, which helps to easily
determine the maximum length of the OneWire without any further instrumentation
{[}2{]}: the most demanding operation of an OneWire bus is the device
discovery procedure. Therefore if for some sensors the discovery does
not reliably work, the capacity of the cable network is in the critical
region and the length should be reduced (reduce number of sensors,
cables or distributors). Even though a normal sensor readout does
still work as it produces a lighter load on the network. A more sophisticated
method is to determine the recovery time of the OneWire Bus {[}3{]}
either with a software implementation on the Arduino (\textquotedblleft pulsein\textquotedblright{}
function) or with an oscilloscope. To comply with the standards, it
should be well above 50\,$\mu\text{s}$, when using not more than
15 sensors at room temperature.

\section{Software }

Although having the full flexibility of an open hardware design, the
software on both the micro controller and collecting server is easy
to deploy. Firstly, because it can be written as a widely-known Arduino
program \cite{key-ArduinoSourcecode} and secondly, the Arduino program
can be flashed online via the Raspberry Pi \cite{key-avrdude-With-RaspberryPi}.

\subsection{Microprocessor Programming }

Due to the broad range of possible sensors, the program of the Arduino
depends strongly on their choice. In the following, we describe the
general structure of the firmware in our setup. After powering on
a sensor the initialization procedure is started. From that point
on, with a frequency of 1\,Hz the information of all sensors is collected
and handed over to the Raspberry Pi. The complete selection of software
used on the Arduino is available online at {[}6{]}.

\subsection{Computer Programs Recommendation for reliable Usage }

In our setup, we are using the Ganglia system \cite{key-1} for the
monitoring of the SAUs and for data collection. Another possibility
is the use of EPICS \cite{key-EPICS} for that purpose, which has
been deployed at another site. Both solutions are capable of collecting
a number of data points in the order of $10^{3}$ on a single computer,
i.e. their visualization and a historical charting functionality.
The data collection server constantly monitors the SAUs and ensures
the full sensor network is in a reliable state. Otherwise it will
issue a software reset on the sensor device or performs a power cycle
via the PoE functionality of the switch.

\appendix

\section*{Acknowledgements}

The authors would like to express their sincere thanks the mechanical
and electronics workshops at the Institut f�r Kernphysik for their
excellent work, which has contributed significantly to the success
of this project. Furthermore, we are grateful the HPC department of
the JGU, for invaluable comments in the design phase.

\end{document}